\documentclass[aps,twocolumn,showpacs,prl,superscriptaddress,unsortedaddress]{revtex4}
\usepackage{epsf}
\usepackage{graphicx}
\usepackage{tabularx, booktabs, makecell, amsmath}
\usepackage{dcolumn}
\usepackage{bm}

\newcommand{\TiSe}{$1T$-TiSe$_2$}

\newcommand{\Tcdw}{$\it{T}_{\text{CDW}}$}

\begin{document}
 \title{ Evidence for breathing-type pseudo Jahn-Teller distortions in the charge density wave phase of \TiSe } 
\author{A. Wegner}
\affiliation{Department of Physics, University of Virginia, Charlottesville, Virginia 22904}
\author{J. Zhao}
\affiliation{Department of Physics, University of Virginia, Charlottesville, Virginia 22904}
\author{J. Li}
\affiliation{Department of Physics, University of Virginia, Charlottesville, Virginia 22904}
\author{J. Yang}
\affiliation{Department of Physics, Central Michigan University, Mount Pleasant, Michigan 48859 22904}
\author{A. A. Anikin}
\affiliation{Department of Physics, Drexel University, 3141 Chestnut Street, Philadelphia, Pennsylvania 19104.}
\author{G. Karapetrov}
\affiliation{Department of Physics, Drexel University, 3141 Chestnut Street, Philadelphia, Pennsylvania 19104.}
\author{D. Louca}\thanks{dl4f@virginia.edu (D.L.)}
\affiliation{Department of Physics, University of Virginia, Charlottesville, Virginia 22904}
\author{U. Chatterjee}\thanks{uc5j@virginia.edu (U.C.)}
\affiliation{Department of Physics, University of Virginia, Charlottesville, Virginia 22904}

\date{\today}

\begin{abstract}
The charge density wave (CDW) phase in \TiSe\ is investigated using angle resolved photoemission spectroscopy (ARPES) and neutron scattering measurements. Our ARPES results reveal a clear temperature dependence of the chemical potential of the system. They also demonstrate specific changes encountered by the Se $4p$ valence and Ti $3d$ conduction bands as the temperature of the system is decreased through \Tcdw. The valence band undergoes a downward shift, whereas the conduction band remains unaffected. The crystal structure in the CDW state shows a distinct split of the Ti-Se atomic correlations that are reminiscent of Jahn-Teller distortions, manifested in a breathing-type mode. The ARPES data together with the local structure analysis support a direct link between Jahn-Teller-like distortions and the CDW order in \TiSe.

\end{abstract}

\maketitle
 It is commonly perceived that the emergent properties observed in most modern day materials, such as cuprate \cite{CUPRATE_REFERENCE1} and pnictide \cite{PNICTIDE_REFERENCE1} high temperature superconductors, colossal magnetoresistive manganites  \cite{CMR_REFERENCE2} and heavy fermion compounds \cite{HEAVY_FERMION_REFERENCE1} arise from competition between or coexistence of proximate broken-symmetry phases.  A unified understanding of the phase competition/coexistence in these strongly correlated systems is, however, intractable due to their abundant disorders and inherent complexity. 
 \TiSe, a transition metal dichalcogenide \cite{TMD_REFERENCE1,TMD_REFERENCE3} CDW compound, is a model system to study because electronic correlations are relatively weak. Moreover, \TiSe\ becomes superconducting in a field-effect transistor configuration  \cite{SC_FET}, via chemical intercalation \cite{CAVA_NP} or under pressure \cite{TISE2_PRESSURE} and is ideal for examining the interplay of multiple degrees of freedom. This is due to the prototypical nature of its CDW order, and its relatively simple crystalline (Fig. 1a) and electronic structures. A critical step towards this quest is to first unveil the origin of the CDW order in the parent compound of \TiSe, an intriguing puzzle despite decades of research.
 
  Bulk \TiSe\ undergoes a CDW phase transition with a 2$\times$2$\times$2 superlattice below the transition temperature \Tcdw$\sim$200 K \cite{MONCTON_NS}. 
  The appearance of superconductivity via chemical intercalation \cite{CAVA_NP} or pressure \cite{TISE2_PRESSURE} within characteristic dome shaped regions of the resultant phase diagrams resemble those of high temperature superconductors and heavy fermion compounds. The intertwining of superconductivity and CDW in the compounds based on \TiSe, however, remains controversial \cite{TiSe2_FENG, TiSe2_HASAN, CAVA_NP}. Recent observations of a robust CDW order in ultra-thin samples \cite{T_C_CHIANG_NC, TAKAHASHI_APL}, and the chiral nature \cite{JSPER_CHIRAL, STM_CHIRAL} of the CDW phase in bulk \TiSe\ have further intensified the search for the precise nature of its CDW transition.

\begin{figure*}[t]
\includegraphics[width=\textwidth]{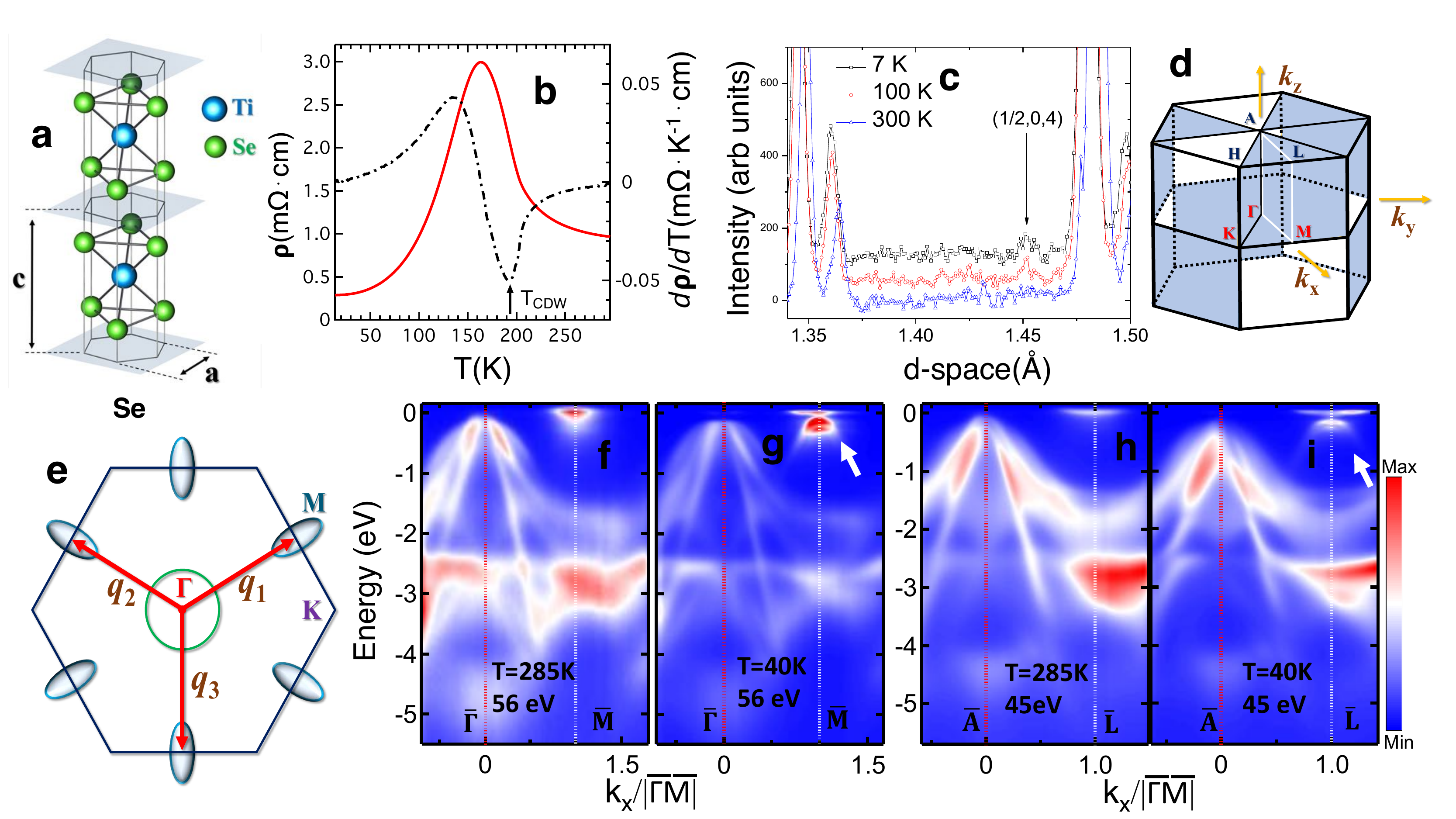}
\caption{(a) Crystal structure of \TiSe. (b) Temperature ($T$) dependence of the in-plane electrical resistivity ($\rho$) of \TiSe. \Tcdw\ is is determined from the minimum (pointed by the black arrow) of  $\frac{d\rho}{dT}$ vs $T$ plot (black dashed line).  (c) The (1/2, 0, 4) Bragg peak in the neutron scattering data corresponds to the CDW superlattice. Its intensity below \Tcdw\ diminishes with increasing temperature and it vanishes above \Tcdw. (d) Schematic layout of the normal state Brillouin zone of \TiSe.  (e) Schematic plot illustrates connections among various sections of the Fermi surfaces via CDW wave vectors $\bf{q}_1$, $\bf{q}_2$ and $\bf{q}_3$. (f,g) EMIMs with $h\nu=56$$\ $eV at $T=285$$\ $K and $T=40$$\ $K, respectively. (h,i) Same as (f,g) but with $h\nu=$45$\ $eV. The red dotted lines correspond to $\overline{\Gamma}$ in (f) and (g), while to $\overline{\text{A}}$ in (h) and (i). The white dotted lines refer to $\overline{\text{M}}$ in (f) and (g), while to $\overline{\text{L}}$ in (h) and (i). The white arrows in (g) and (i) indicate the backfolding of Se $4p$ bands below \Tcdw\ at $\overline{\text{M}}$ and $\overline{\text{L}}$, respectively. }
\end{figure*}

Similarly to a number of other CDW bearing TMDs \cite{UC_NC, UC_PRB, UC_JMC}, the Fermi surface nesting scenario \cite{CDW_GRUNNER} does not apply to \TiSe. Existing models of the CDW order present diverging explanations and can be broadly grouped into  three categories based on the possible roles of the lattice and electronic degrees of freedom: (i) In an excitonic condensation mechanism \cite{EXCITONIC_INSULATOR_2}, excitons above \Tcdw\ are formed because of poorly screened Coulomb interaction and the long-range CDW order emerges due to Bose-Einstein condensation (BEC) of excitons below \Tcdw. (ii) In the case of a Jahn-Teller (JT)-like distortion mechanism \cite{JT_HUGE, JT_WANGBO}, mediated by electron-phonon interactions, the CDW transition occurs through a partial charge transfer between the Se $4p$ valence and Ti $3d$ conduction bands. Two different types of distortions have been speculated. One of them is a band type JT-like distortion (BJT) \cite{JT_HUGE} proposed by Hughes that corresponds to a relative rotation between the neighboring octahedra consisting of Se atoms inside the \TiSe\ lattice. The other is the pseudo JT-like distortion (PJT) proposed by Whangbo and Canadell \cite{JT_WANGBO}, where a shortening of the Ti-Se bonds stabilizes the structural distortion below \Tcdw. (iii) A hybrid model \cite{JASPER_EPL_ALTERNATIVE} suggests a cooperative combination of JT-like distortions and exciton condensation to be the origin of the CDW transition. 

Some of the earlier ARPES measurements have been interpreted in the framework of excitonic condensation  below \Tcdw\ \cite{AEBI_1, AEBI_2}. The BEC of excitons below \Tcdw\ has also been suggested from momentum-resolved electron energy loss spectroscopy measurements \cite{PETER_SCIENCE}. The detection of the CDW correlations even after non-thermal melting of the excitonic insulating phase via ultrafast optical spectroscopy \cite{KYLE_ULTRAFAST_OPTICS}, however, is incompatible with a purely electronic origin  of the CDW transition. Additionally, recent scanning tunneling microscopy (STM) measurements \cite{AEBI_STM2} and first principles calculations \cite{TISE2_DFT1, TISE2_DFT2} emphasized the significance of local distortions to the CDW instability in \TiSe. Collectively, these observations call for a direct investigation of atomic scale lattice distortions and their influence on the electronic structure of \TiSe\ in the CDW state. Note that previous structural measurements have mostly looked into softening of phonon modes \cite{HOLT_PRL, FRANK_PRL}, but a direct investigation of the local structural distortions, essential to the JT mechanism, is lacking. 

In this work, we study the temperature-dependent structural and electronic properties of \TiSe\ employing neutron scattering and ARPES measurements. This enables us to identify not only the lattice distortions at the atomic scale but also their impact on the energy bands. Our results establish for the first time the following: (i) a local structural distortion, realized in Ti-Se bond shortening giving rise to a breathing mode, in the CDW state; (ii)  temperature dependence of the chemical potential $\mu$, of the system; and (iii) a temperature-dependent downward shift of the Se $4p$ band, while leaving the Ti $3d$ band intact. We emphasize that it is the observation of the temperature dependence of $\mu$ in the current work, which enables us to discern the aforementioned dichotomy in the temperature evolution of the valence and conduction bands. 

We carried out temperature-dependent ARPES measurements on \TiSe\ single crystals using a Scienta R4000 electron analyzer at the PGM beamline of the Synchrotron Radiation Center, Stoughton, Wisconsin, USA  and  the SIS beamline of Swiss Light Source, Paul Scherrer Institute, Switzerland.  The ARPES measurements were performed using plane polarized light with 45 and 56 eV photon energy ($h\nu$). The energy and momentum resolutions were approximately 10-20 meV and 0.0055 $\AA^{-1}$  respectively. Single crystals were cleaved $\it {in}$ ${situ}$ to expose a fresh surface of the crystal for ARPES measurements. The neutron scattering experiments were carried out using the BT-1 diffractometer at the NIST Center for Neutron Research and at the Nanoscale Ordered Materials Diffractometer (NOMAD), Oak Ridge National Laboratory. The neutron scattering data were analyzed both in reciprocal and real-space to obtain the average and local structures, respectively. Scattering from the sample container and dixplex were subtracted from the NOMAD data and corrected for absorption and multiple scattering and normalized by vanadium. The resulting total structure factor, S(Q), was Fourier transformed in real-space to obtain the pair distribution function, G(r). A maximum momentum transfer, $Q_{max}=40$ \AA $^{-1}$, was used. The G(r) corresponds to a real-space representation of  the local atomic correlations. 

The combined characterization of the CDW transition in \TiSe\ via neutron scattering, electrical transport and ARPES  measurements is displayed in Fig. 1. \Tcdw$\sim$200$\ $K can be determined from the onset of the upsurge in the electrical resistivity vs temperature curve in Fig. 1(b). The system undergoes a structural transition at \Tcdw\ from an undistorted $P\overline{3}m1$ symmetry to a 2$\times$2$\times$2 superlattice with $P\overline{3}c1$ symmetry, giving rise to half integer Bragg peaks for temperatures below \Tcdw\   \cite{CDW_GRUNNER}. One such peak is indicated in Fig. 1(c). Moreover, the CDW leads to a zone folding and thus, back-folded CDW energy bands \cite{CDW_GRUNNER} are anticipated in the momentum space (Figs. 1(d), 1(e)). We inspect this by focussing on ARPES energy-momentum intensity maps (EMIM's) in Figs. 1(f)-1(i). An EMIM is ARPES data as a function of one of the in-plane momentum components ($k_x$ in the current case) and electronic energy $\overline{\omega}$ referenced to $\mu$ for a fixed value of $k_y$. 

We describe the EMIMs in terms of the surface Brillouin zone (BZ) high-symmetry points. The normal emission ARPES spectra with $h\nu=$ 56$\ $eV and 45$\ $eV correspond to the states located in the vicinity of $\Gamma$ and A points respectively (Fig. 1(d)). Therefore, the high symmetry points of Figs. 1(f) and (g) are referred by $\overline{\Gamma}$ and $\overline{\text{M}}$, while those of Figs. 1(h) and (i) by $\overline{\text{A}}$ and $\overline{\text{L}}$. The energy bands closest to $\mu$  around $\overline{\Gamma}$ and $\overline{\text{A}}$ are due to the Se $4p$ bands, while those around $\overline{\text{M}}$  and $\overline{L}$ correspond to the Ti $3d$ bands. The formation of a 2$\times$2$\times$2 CDW order leads to the backfolding of the Se 4p bands at $\overline{L}$ and $\overline{\text{M}}$ points below \Tcdw. This can be observed from Figs. 1(g) and 1(i), which correspond to temperature $\sim$40$\ $K$<$\Tcdw. Moreover, the backfolded bands are practically non-existent in Figs. 1(f) and (h), which correspond to a temperature $\sim 285$$\ $K$>$\Tcdw.  Some residual intensity from the backfolded bands is, however, visible  above \Tcdw. This would imply short-range CDW correlations above \Tcdw. These results agree well with previous ARPES studies on \TiSe\ \cite{TiSe2_FENG, TiSe2_HASAN, AEBI_1, AEBI_2, PILLO_PRB, ROSSNAGEL_PRB}.

%\begin{figure*}[ht!]
\begin{figure}[h!]
\centering
\includegraphics[width=0.45\textwidth]{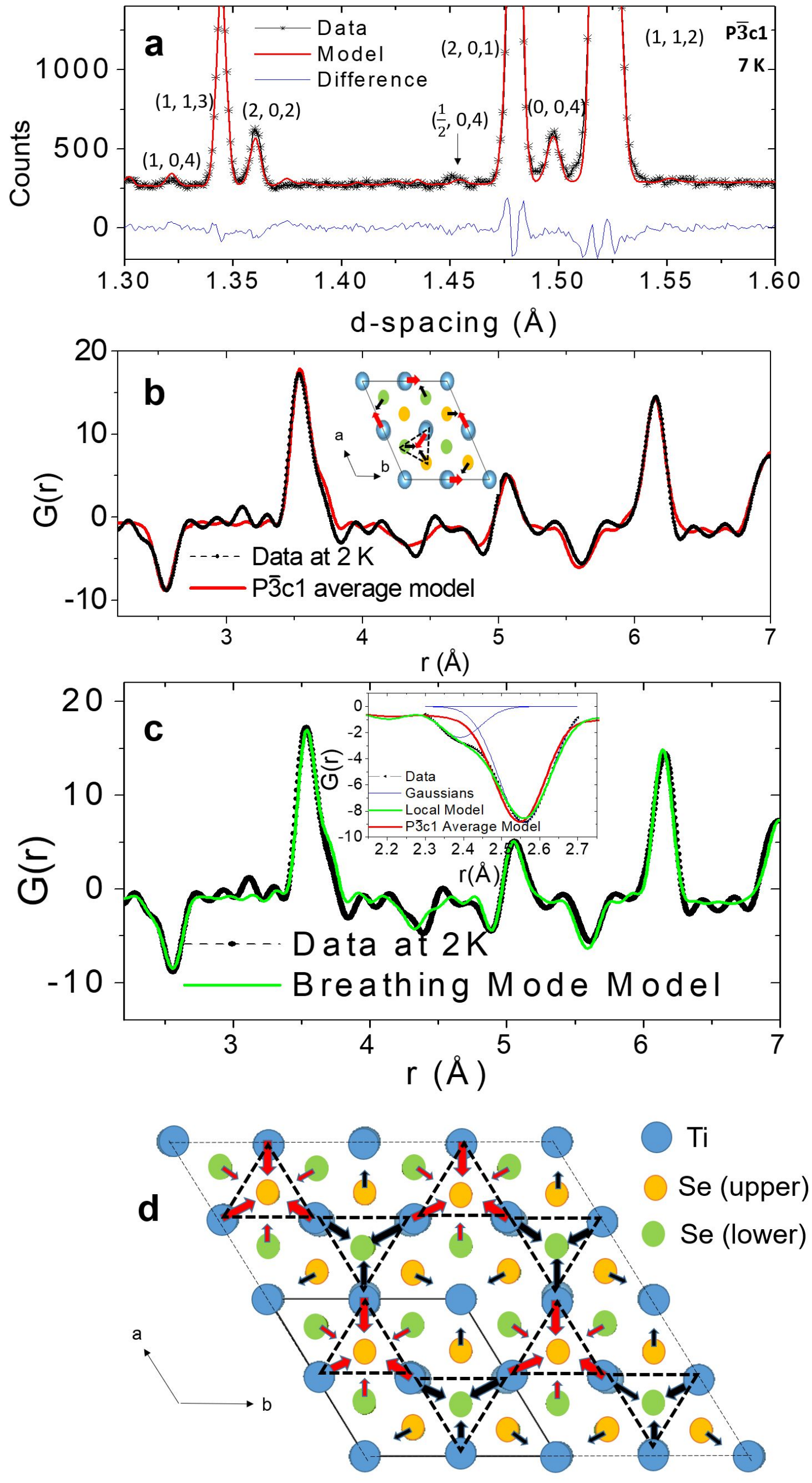}
\caption{(a) The neutron diffraction data at 7 K is compared to the model calculated from $P\overline{3}c1$ symmetry. The superlattice peak associated with the CDW is reproduced with the 2$\times$2$\times$2 supercell at 7 K. The difference is shown in blue. (b) The local structure is compared to the average structure model first reported by Di Salvo \cite{MONCTON_NS}. The average structure model does not reproduce the splitting of the first nearest Ti-Se bonds. Distortions associated with the model are shown in the inset.  (c) A model that includes a breathing mode fits the local structure including the first peak splitting. A close-up of the first peak including the breathing model and a two Gaussian fit of the peak is displayed in the inset. (d) An expanded view of four superlattice unit cells showing the breathing mode motif, indicated by arrows. The clusters of shortened Ti-Se bonds are indicated by the dashed triangles.}
%\end{figure*}
\end{figure}

To carefully examine the structural motifs of the CDW phase, we turn our attention to Fig. 2.  Superlattice reflection, indicated by the arrow in the plotted range, arise from the 2$\times$2$\times$2 lattice expansion. The diffraction data in Fig. 2(a) are fit to a model calculation based on the low-temperature $P\overline{3}c1$ hexagonal symmetry of the average structure, with $a=b=7.05232\pm0.00008$ \AA\ and $c=11.9816\pm 0.0003$ \AA. First proposed by DiSalvo et al. \cite{MONCTON_NS}, the CDW structural modulations in the hexagonal unit cell are shown in the inset of Fig. 2(b). In this model, the Ti atoms are displaced towards other Ti atoms and the magnitude of the displacement is 0.085 \AA\ while the displacement of the Se atoms is 0.028 \AA\ toward other Se atoms. Fig. 2(b) is a plot of the G(r) obtained from the data collected at 2 K (symbols) from NOMAD. The first peak in the G(r) at ~2.55 \AA\ corresponds to the shortest bond length in the crystal structure involving pairs of Ti-Se atoms in the triangle indicated in the inset. The peak is negative because Ti has a negative neutron scattering length \cite{SCATTERING_LENGTH}. Also shown in this figure is a model G(r) calculated based on the $P\overline{3}c1$ CDW phase. While overall this symmetry fits the data at 2K well with a $\chi ^{2}$ $\sim $ 0.592, it fails to reproduce the split of the Ti-Se correlations as seen in the inset of Fig. 2(c). The split of the Ti-Se
correlation peak gives rise to a distribution of long and short bonds centered at $\sim$2.39 \AA\ and 2.55 \AA. On the other hand, in the $P\bar{3}c1$ model, the Ti-Se nearest neighbor bond lengths range from 2.50 to 2.57 \AA. The change in bond lengths assumed in the DiSalvo model is not sufficiently large to explain the 0.16 \AA\ peak split observed in the local atomic structure. A better fit to the data is provided by a local model that involves breathing type distortions of Ti
toward Se atoms above and below the Ti plane as shown in Fig. 2(d). To reproduce the split of the Ti-Se pair correlation peak, the restrictions on the Ti-Se motion is relaxed from the $P\overline{3}c1$ symmetry and the Ti atoms are constrained to move toward Se atoms rather than toward other Ti atoms as in the DiSalvo model. The local model is constructed using the $P\overline{3}m1$ symmetry of the system above \Tcdw\ and the parameters are provided in the supplemental section. The model G(r) calculated from the breathing type distortion is compared to the data in Fig. 2(c). While the overall agreement is quite good, this model
particularly fits the Ti-Se pair correlations as seen in the inset of Fig. 2(c). The Ti distortions in the breathing model are ~ 0.12 \AA\ that results in Ti-Se bond lengths ranging from 2.41 to 2.57 \AA. The $\chi^{2}$ of this fit is 0.252.

In the case of the BJT-like distortions \cite{ JT_HUGE}, the top and bottom faces of the octahedra comprised of Se atoms, and surrounding Ti atoms at the corners of the unit cell in the CDW state, would rotate in opposite directions. This is not consistent with our G(r) data. On the other hand, in case of the PJT-like distortions, the Ti atoms must move to shorten some of the Ti-Se bonds to stabilize the distortion. This agrees with our breathing model, where the large splitting of the Ti-Se peak observed in the G(r) may arise from a JT-like mechanism. These imply that  the Ti-Se bond-shortening is crucial to the stabilization of the lattice distortion of \TiSe\ below \Tcdw, which was also conjectured in a recent STM work \cite{AEBI_STM2}.

\begin{figure}[h]
\includegraphics[width=3.4in]{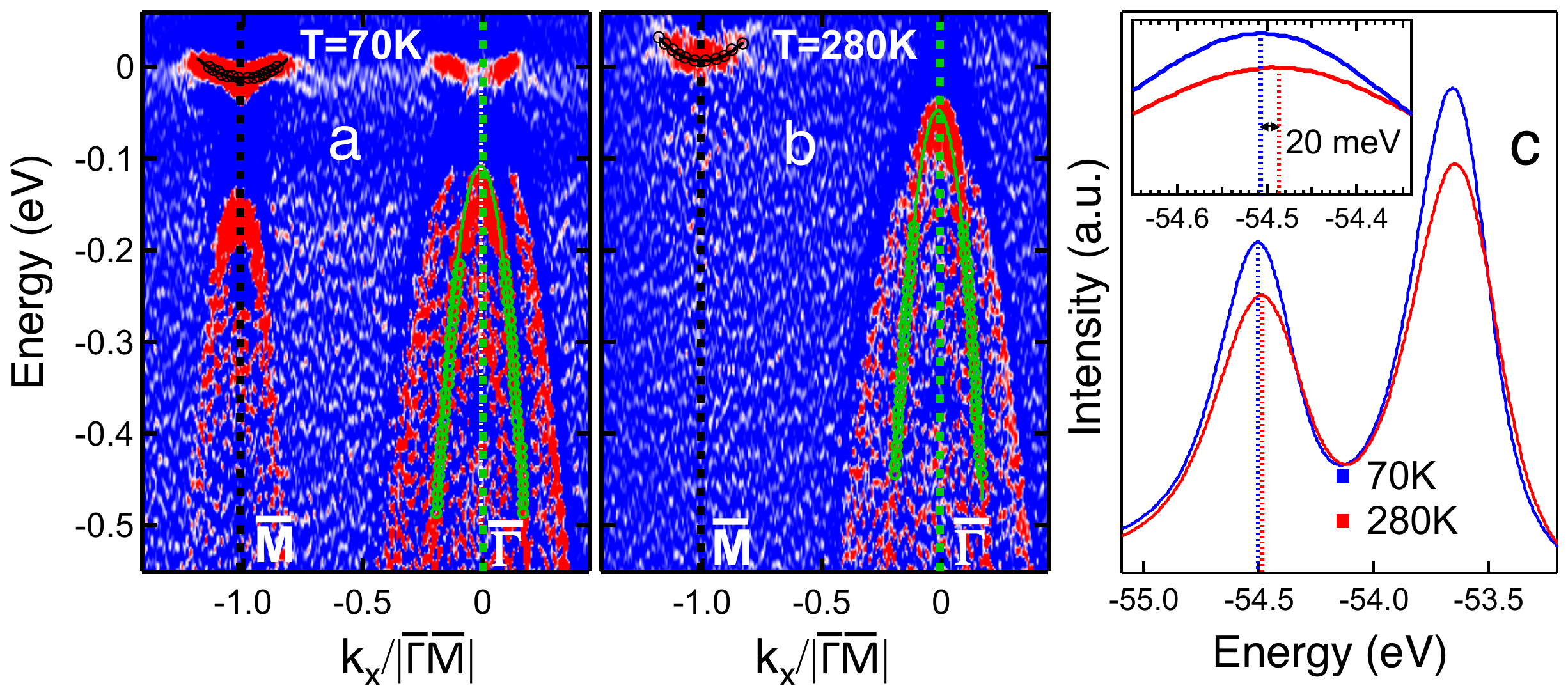}
\caption{(a) Second derivative (with respect to $\overline{\omega}$) of an EMIM, taken with $h\nu=$56 eV, at $T=$70$\ $K. (b) Same as (a) but at $T=$280$\ $K. 
The green and black dotted lines in (a) and (b) correspond to $\overline{\Gamma}$ and $\overline{\text{M}}$. Dispersions at $\overline{\text{M}}$ (black open circles) are fitted with with parabolic curves (black solid lines) at both low and high temperatures.  Similar fittings (green solid lines) have also been done for the dispersions (green open circles) at $\overline{\Gamma}$. 
(c) Core level spectra of Se with a focus on the Se $3d_{\frac{3}{2}}$ and $3d_{\frac{5}{2}}$ peaks at  $T=$70$\ $K and 280$\ $K. Inset shows an expanded view of the Se $3d_{\frac{3}{2}}$ peaks at 70$\ $K and 280$\ $K.}
\end{figure}

What is the impact of the above-described JT-like distortions on the electronic structure of \TiSe? In both the BJT and PJT models  \cite{JT_HUGE, JT_WANGBO},
an exchange of single-particle spectral weights between the Se $4p$ valence and Ti $3d$ conduction bands is responsible for the net reduction in the free energy of the CDW state over the normal state. With decreasing temperature through \Tcdw, the energy lowering for the BJT and PJT cases occurs via a downward movement of the Ti $3d$  and Se $4p$ bands, respectively. In order to verify this, we compare the temperature-dependent changes in the band dispersion of the valence and conduction bands in Fig. 3. Dispersions of the valence and conduction bands at different temperatures are detected using second derivative analysis. Figs. 3(a) and 3(b) correspond to the second derivatives of EMIMs around $\overline{\Gamma}$$ \overline{\text{M}}$ at temperatures of $70$$\ $K and $280$$\ $K, respectively. Apparently, both the valence and conduction bands shift towards a higher binding energy as temperature is decreased from 280$\ $K ($>$\Tcdw) to 70$\ $K($<$\Tcdw). The magnitude of the shift of the valence band is, however, substantially larger than that of the conduction band. From this, like in previous ARPES studies \cite{AEBI_1, AEBI_2, PILLO_PRB, ROSSNAGEL_PRB}, it would seem that differentiation between the PJT and BJT scenarios based on ARPES data alone is not feasible. However, this is not the case provided the temperature dependence of $\mu$ is taken into account. The temperature dependence of $\mu$ can be approximated from the core level peaks \cite{PHOTOEMISSION_BOOK_HUFFNER}. Here, we monitor the Se $3d_{\frac{3}{2}}$ and $3d_{\frac{5}{2}}$ core level peaks at $T=280$K and $T=70K$ in Fig. 3(c), the inset of which illustrates that like the Ti $3d$ and Se $4p$ bands, the $3d_{\frac{3}{2}}$ core level peak also undergoes a shift to higher binding energies as temperature is lowered from 280 to 70$\ $K. Strikingly, this shift in the core level spectra is the same as that of the conduction band within experimental error. This leads us to conclude that the temperature-dependent shift of the band minima of the Ti $3d$ bands is rooted to $\mu(T)$, while the evolution of the Se $4p$ band with temperature is indeed nontrivial.  

To summarize, using our neutron scattering measurements, we report the first observation of JT-like distortions in the CDW state of \TiSe. From our ARPES data, we highlight strikingly distinct temperature evolutions of the valence and conductions bands. In sharp contrast to the valence band that shifts towards a higher binding energy with decreasing temperature, the conduction band remains unchanged. With regard to the driving mechanism of the CDW instability in \TiSe, these results are in favor of the hypothesis that the PJT-type distortions trigger the CDW instability.

UC acknowledges supports from the National Science Foundation (NSF) under Grant No. DMR- 1629237 and from the Jefferson Trust at the University of Virginia. DL  acknowledges support from the Department of energy, contract No.  DE-FG02-01ER45927. GK and AAA acknowledges support from the NSF under Grant No. ECCS-1711015. UC thankfully acknowledges fruitful discussions with J. van Wezel.

\end{document}